\documentclass[prl,twocolumn,showpacs]{revtex4}
\usepackage{amsmath}
\usepackage[dvips,xdvi]{graphicx}
\usepackage{amssymb}
\usepackage{epsfig}

\begin{document}
\title{Quantum Phase Transitions and Continuous Observation of Spinor Dynamics in an Antiferromagnetic Condensate}
\author{Y. Liu$^1$}
\email{yingmei.liu@nist.gov}
\author{S. Jung$^1$}
\author{S. E. Maxwell$^1$}
\author{L. D. Turner$^2$}
\author{E. Tiesinga$^1$}
\author{P. D. Lett$^1$}

\affiliation{$^1$Joint Quantum Institute, National Institute of
Standards and Technology and \\University of Maryland,
Gaithersburg, Maryland 20899\\$^2$School of Physics, Monash
University, Victoria 3800, Australia}
\date{\today}

\begin{abstract}
Condensates of spin-1 sodium display rich spin dynamics due to the
antiferromagnetic nature of the interactions in this system. We
use Faraday rotation spectroscopy to make a continuous and
minimally destructive measurement of the dynamics over multiple
spin oscillations on a single evolving condensate. This method
provides a sharp signature to locate a magnetically tuned
separatrix in phase space which depends on the net magnetization.
We also observe a phase transition from a two- to a
three-component condensate at a low but finite temperature using a
Stern-Gerlach imaging technique. This transition should be
preserved as a zero-temperature quantum phase transition.

\end{abstract}

\maketitle
The study of multi-component Bose-condensed
(superfluid) systems began with $^4$He - $^6$He mixtures in the
1950's~\cite{he4} and continued with a two-component $^{87}$Rb
Bose-Einstein condensate (BEC), with atoms in two different
hyperfine states, in the late 1990's~\cite{mBEC}. Recent studies
have focused on the investigation of spinor condensates where
interconversion among multiple spin states leads to spin
population dynamics. A number of investigations of this effect, as
well as spatial domain formation in both ferromagnetic $F$=1
$^{87}$Rb BECs~\cite{chapman1,chapmanQ,lyou1} and
antiferromagnetic $F$=1 $^{23}$Na
BECs~\cite{ketterle1,ketterle2,lett} have been published. The
$F$=2 $^{87}$Rb spinor condensate presents ferromagnetic,
antiferromagnetic, and cyclic
phases~\cite{bloch,f2Hirano,f2sengstock1,f2sengstock2}. In each of
these cases the experimental system can be modelled with a small
number of variables.

While the ferromagnetic Rb system is becoming well-studied, the
antiferromagnetic Na system remains relatively unexplored. The
theoretical description of such a system suggests the possibility
of manipulating the phase space topology and dynamics of the
system in ways not possible in the ferromagnetic system (for
example, altering the separatrix position in phase space with the
magnetization of the system) as well as the possibility of
observing a quantum-fluctuation-driven phase transition that does
not exist in the ferromagnetic system~\cite{lyou1,gs}.

We consider a spinor BEC with spin angular momentum $F=1$ in the
presence of a magnetic field of strength $B$ along the $z$ axis
with the populations initialized to a non-equilibrium state.
Collisional interconversion between two $m_{F} = 0$ atoms and one
$m_{F} = +1$ and one $m_{F} = -1$ atom takes place in the
condensate, leading to oscillations in the spin populations. At
ultracold temperatures the collisions between alkali metal atoms
conserve the summed spin angular momentum $\vec f=\vec F_a + \vec
F_b$. Our system, $^{23}$Na, is antiferromagnetic, or polar,
inasmuch as the interaction energy of $f=2$ collisions is larger
than that of $f=0$ collisions, which indicates that the coupling
favors the $m_{F}=\pm1$ states over the $m_{F} = 0$ state.

The linear Zeeman shift induced by the magnetic field does not
affect the collisional interconversion, as the magnetic energies
before and after the collision are equal in this approximation.
The population dynamics are instead driven by an interplay between
the quadratic Zeeman shift and the spin-dependent interaction
characterized by the difference in the $f=0$ and $f=2$ interaction
energies. In $^{23}$Na, a divergence in the spin oscillation
period occurs near a critical magnetic field
$B_c$~\cite{lyou1,f2sengstock2,lett}. A dependence of $B_c$ on
magnetization $m$ (the difference in fractional population
$\rho_{m_F}$ between $m_{F} = +1$ and $m_{F} = -1$) is predicted
in an antiferromagnetic system~\cite{lyou1} or in a ferromagnetic
system with a radio-frequency (rf) dressing
field~\cite{dressstate}, but has not been previously observed.

We use two complementary methods to observe the spin dynamics in
two different time regimes. A Stern-Gerlach separation followed by
absorption imaging (SG-AI) is the standard method to directly
measure the populations of different spin states and determine the
magnetization $m$. This technique is, however, completely
destructive and only minimal phase information about the spin
oscillations can be inferred from modelling the data. The second
method is Faraday rotation spectroscopy which measures the
rotation of the polarization of a laser beam. This rotation is
proportional to the projection of the atomic spin $\vec F$ along
the laser propagation direction. It can be used to continuously
infer both phase and population information of the spin dynamics
over multiple spin oscillations. Other methods of measuring the
condensate phase can be found in~\cite{phase1}.

It is hard to determine $B_c$ from just the spin oscillation
period, however, we observe a sharp signature to distinguish two
characteristic time evolutions in the vicinity of $B_c$ with
Faraday rotation spectroscopy. At long times when the oscillations
have damped out~\cite{lett}, we use SG-AI to characterize the
mean-field ground state populations as a function of $B$ and $m$
to observe a predicted phase transition from a two- to a
three-component spinor BEC.

A $^{23}$Na magneto-optical trap containing up to $6\times10^9$
atoms is prepared. A crossed optical dipole trap derived from a
multi-mode laser at 1070~nm is then loaded, followed by
evaporation and re-thermalization. A weak magnetic field gradient
is applied during 6 s forced evaporation to form a fully polarized
BEC of $1.5\times10^5$ atoms in the $| F=1, m_F = 1 \rangle$
state. The final trap frequencies are $\omega_{x,y,z} \approx (2
\pi) 220 (\sqrt{2},1,1)$~Hz, and the mean Thomas-Fermi radius is
7.2 $\mu$m. We ramp up the magnetic field along the $z$ axis while
turning off the field gradient. The final value of $B$ ranges
between 6.3 $\mu$T and 60.7 $\mu$T with an uncertainty of 0.04
$\mu$T (uncertainties in this Letter represent estimated one
standard deviation combined statistical and systematic
uncertainties). We can prepare an initial state with any desired
$m$ and $\rho_0$ by an rf-pulse to rotate the atomic spin followed
by selective removal of atoms in a given $m_F$ state. The rf-pulse
is resonant with the linear Zeeman splitting, and its amplitude
and duration control the superposition of the $m_F$ levels. The
removal is performed by a microwave pulse to selectively transfer
$|F=1, m_F\rangle$ atoms to the $F=2$ state, followed by a laser
pulse resonant with these atoms.

\begin{figure}[htbp]
\includegraphics[width=80mm]{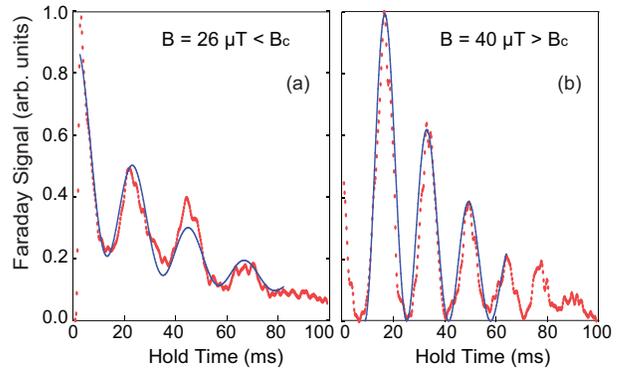}
\caption{Faraday signal (proportional to $\langle F_x\rangle^2$)
taken from a single measurement for $m=0$ at two magnetic fields,
26 $\mu$T and 40 $\mu$T starting with $\rho_0=0.5$, $\theta=0$.
The solid line is a fit with a damped sinusoid. The signals show
an oscillating phase (a) and a running phase (b) at $B$ below and
above $B_c$, respectively, as evidenced by the signal reaching
zero or not. } \label{Faraday}
\end{figure}

The Faraday detection beam is directed along the $x$ axis and
red-detuned 225~GHz from the 10 MHz wide D2 line of $^{23}$Na. The
beam is linearly polarized, has a $1/e^2$-waist of 1~mm at the
condensate, and a power of $\approx$50 mW. The set-up for Faraday
spectroscopy is similar to that outlined
in~\cite{fsJessen1,fatemi}. A carefully chosen aperture is
inserted into the imaging plane for an optimal signal-to-noise
ratio (SNR), and the Faraday rotation of the linear polarization
is detected using a Wollaston prism and an autobalanced
differential photodetector (PD)~\cite{hobbs}. The Faraday rotation
angle oscillates at the Larmor precession frequency $f_L$. Changes
of spin populations and phases are detected as a modulation of the
amplitude of the Faraday signal. Our Faraday signal is the
short-time power spectral density of the PD output integrated over
a narrow bandwidth of 1 kHz around $f_L$. This is proportional to
the slowly varying envelope of $\langle F_x\rangle^2$. A typical
example is shown in Fig.~\ref{Faraday}. We divide the Faraday
signal into 1 ms time bins, longer than the transform limit of the
digital filter (0.16 ms) but short enough to resolve spin
oscillations. Over a 100 ms measurement, we thus make 100 distinct
measurements of both the spin projection amplitude and $f_L$, on a
single evolving spinor BEC.

The parameters of the Faraday beam are chosen to minimize atom
loss from the off-resonant light scattering while maintaining a
good single-shot signal. The measured lifetime of the BEC in the
presence of the Faraday beam is 100 ms, consistent with the decay
of $\langle F_x\rangle$ inferred from Fig.~\ref{Faraday}, and with
the predicted photon scattering time. The dephasing time due to
the tensor light shift~\cite{fsJessen2} is one order of magnitude
longer. The scattering loss is larger than any other back-action
in the present experiments. In the absence of the Faraday beam, we
observe an energy dissipation which depends on $B$ and the mean
particle density $\langle n\rangle$. Under the conditions of
Fig.~\ref{Faraday}(a), the time scale of this dissipation is five
times longer than the decay seen in this figure, while at high
fields (Fig.~\ref{Faraday}(b)), it becomes comparable. This
dissipation is not well understood.

The SNR of our measurements is limited by the total number of
atoms in the BEC and the efficiency of the
detection~\cite{fsJessen1}. Our BECs are not much larger than what
is required to get a good single-shot signal with our system. Our
overall detection efficiency could be improved by a factor of two.

The single mode approximation (SMA)~\cite{lett,lyou1} is applied
to understand our data. The spatial wavefunction of the BEC is
treated as a single mode and the unit-normalized total
wavefunction can be represented as $\Psi(\textbf{r},t) =
\Phi(\textbf{r})(\sqrt{\rho_{-1}(t)}e^{i\theta_{-1}(t)},
\sqrt{\rho_{0}(t)}e^{i\theta_{0}(t)},\sqrt{\rho_{+1}(t)}e^{i\theta_{+1}(t)})$,
where $\rho_{m_F}$ and phases $\theta_{m_F}$ are independent of
position. The Hamiltonian conserves particle number and $m$. The
system is described using $\rho_0$ and $\theta =
\theta_{-1}+\theta_{+1} - 2\theta_0$, with the conserved classical
spinor energy
\begin{equation}
E=E_{\rm qz}(1-\rho_0)+
c~\rho_0\left((1-\rho_0)+\sqrt{(1-\rho_0)^2-m^2}\cos\theta\right).\label{eq1}
\end{equation}
Here $E_{\rm qz}$ is the quadratic Zeeman shift ($E_{\rm qz}/h
=(0.0277 \,\textrm{Hz}/( \mu\textrm{T})^2)B^2$), $c = c_2 \langle
n\rangle $ is the spin-dependent collision energy, and $c_2$ is
$1.59\times10^{-52}$ Jm$^3$ for $^{23}$Na~\cite{lett}. The
evolution of $\rho_0$ and $\theta$ is given by
$\dot{\rho_0}=-(2/\hbar)\partial E /\partial \theta$ and
$\dot{\theta}=(2/\hbar)  \partial E /\partial \rho_0$.

\begin{figure}[htbp]
\includegraphics[width=80mm]{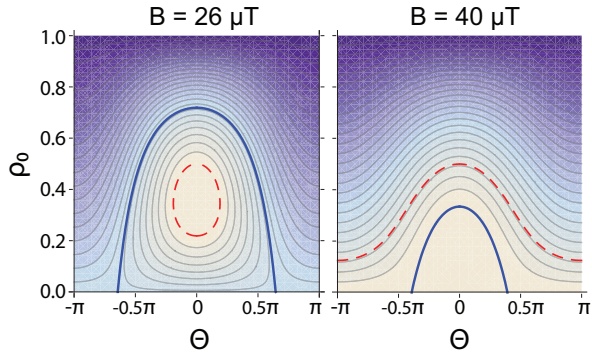}
\caption{(Color online) Equal-energy contour plots generated from
Eq.~\ref{eq1} at two magnetic fields, 26 $\mu$T (left) and 40
$\mu$T (right), with $m=0$ and $c/h = 33 $ Hz. The (red) dashed
lines represent the energy of a state with $\rho_0(t=0)=0.5$. The
heavy (blue) solid lines represent the energy of the separatrix
($E_{\rm sep}$) between oscillating and running phase solutions.
Darker colors represent lower energies.} \label{phasediagram}
\end{figure}

Figure~\ref{phasediagram} shows typical phase diagrams of the
equal energy contours of Eq.~\ref{eq1} for $m=0$ at two magnetic
fields. The preparation of the state determines the energy
$E_0(B)$. Our initial states have $\theta=0$. At any magnetic
field, we can define a separatrix, i.e. that energy contour,
$E_{\rm sep}(B)$, on which there is a saddle point where
$\dot{\rho_0}=\dot{\theta}=0$. This defines the boundary between
two regions in phase space. In fact, $E_{\rm sep}(B)= E_{\rm qz}$
in our system.

When $E_0 (B) > E_{\rm sep}(B)$, the value of $\theta$ is
restricted, while for $E_0(B) < E_{\rm sep}(B)$ there is no bound.
This defines regions with oscillating phase and running phase,
respectively. In both regions $\rho_0$ oscillates. At $E_0(B)\gg
E_{\rm sep}$, which corresponds to small magnetic fields, the
period only weakly depends on the field. In the opposite limit,
the period decreases rapidly with increasing field. In both cases,
the oscillation is nearly harmonic. When $B\sim B_c$, defined by
$E_{0}(B_c)=E_{\rm sep}(B_c)$, the oscillation is anharmonic and
the period diverges for $B=B_c$.

In the SMA, the Faraday signal is derived from
\begin{eqnarray}
\langle F_x\rangle &=&\cos\left[\frac{\theta +(\theta _{+1}-\theta
_{-1})}{2}\right] \sqrt{\rho _0\left(1+m-\rho _0\right)
}\label{eq2}\\
&&+\cos\left[\frac{\theta -(\theta _{+1}-\theta _{-1})}{2}\right]
\sqrt{\rho _0 \left(1-m-\rho _0\right)}\nonumber.
\end{eqnarray}
The phase difference $\theta_{+1}-\theta_{-1}$ is determined by
the fast Larmor precession and a slow evolution due to $\rho_0$
and $\theta$~\cite{lyou1}. For $m=0$, our Faraday signal is
proportional to $\rho_0(1-\rho_0)\cos^2(\theta/2)$. For
oscillating phase solutions, $\theta$ oscillates about zero (with
amplitude $<\pi$) and thus the signal is always greater than zero.
On the other hand, the signal is periodically zero for running
phase solutions. Figure~\ref{Faraday} shows signals from the two
regions. For $m\neq 0$, the signal is described by a more
complicated expression, but the distinction between the two
regions in the vicinity of $B_c$ remains the same.

\begin{figure}[htbp]
\includegraphics[width=80mm]{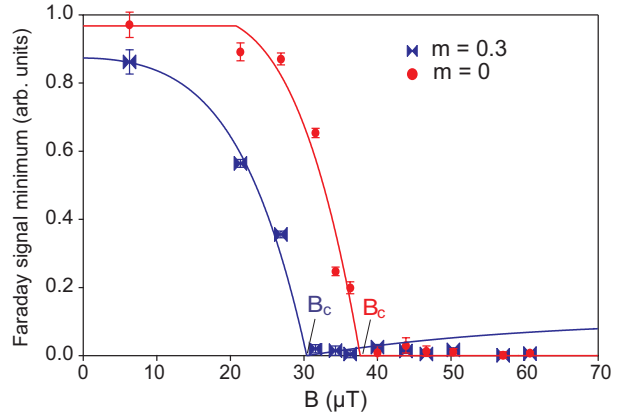}
\caption{(Color online) The minimum of the Faraday signal as a
function of magnetic field for $m=0$ (red dots) and $m=0.3$ (blue
bowties). A scale factor is applied to the Faraday signal to
correct for the PD response at different $f_L$. The lines are fits
based on Eq.~\ref{eq2}, which yield the fit parameters
$\rho_0=0.42$ and $N=1.50\times10^5$ for $m=0$, and $\rho_0=0.54$
and $N=1.32\times10^5$ for $m=0.3$. The fit parameters are within
the 3\% uncertainty of those derived from absorption images.}
\label{diffmag}
\end{figure}

These characteristics of the Faraday signal provide a sharp
signature for locating $B_c$. Figure~\ref{diffmag} shows the value
of the minimum of the Faraday signal at different $B$ and $m$
after removal of the exponential decay. For the two magnetizations
a transition from an oscillating phase solution with a non-zero
minimum to a running phase solution with a minimum of zero,
provides the sharp signature to locate $B_c$. In \cite{lett}, the
population oscillations were measured using SG-AI and fit by a
sinusoid to extract a period and to locate the two regions of
phase space. No sharp experimental signature distinguishing the
boundary was observed.

Figure~\ref{diffmag} also shows a comparison between the
prediction from the SMA and the data. For $m=0$, the agreement is
excellent. For $m=0.3$, however, the prediction does not agree
with our measurements for fields significantly larger than $B_c$.
At the transition point, the minimum of the Faraday signal goes to
zero, but above this point, the theory predicts that the minimum
rises with $B$, even though the solution has a running phase. This
increase is not observed. The apparent agreement between the SMA
theory and our measurements has been surprising - at every
magnetic field reported in this paper, we have seen the presence
of several spatial modes/spin domains during the spin
oscillations, although not in steady-state. The observation of
spin domains is in marked contrast to our previous
work~\cite{lett}. Several technical changes may have contributed
to the domain formation, such as a 50\% increase of the total atom
number and more stable magnetic bias fields. Understanding the
ways in which multiple spatial modes and domains affect the spin
dynamics is an interesting direction for future research.

\begin{figure}[htbp]
\includegraphics[width=80mm]{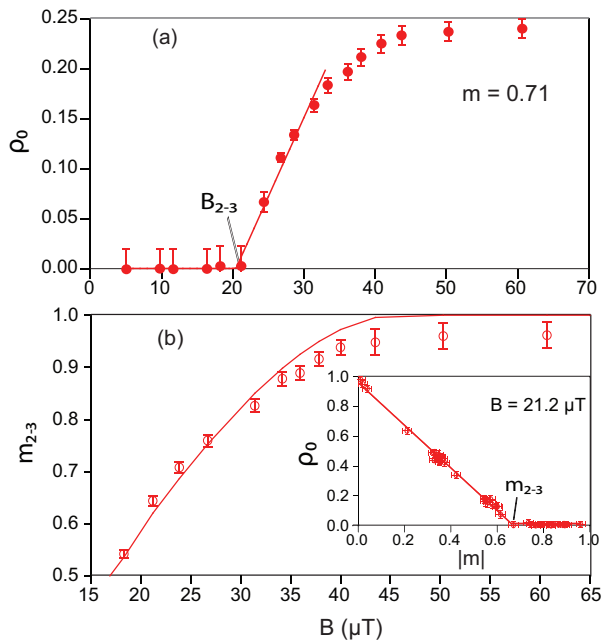}
\caption{Evidence of phase transitions in the mean-field ground
state of the antiferromagnetic spinor BEC. (a): Points with
uncertainties indicate $\rho_0$ as a function of $B$ at $m =
0.71(2)$. The intersection of two linear fits defines $B_{2-3}$,
the critical magnetic field for the phase transition. (b) Inset:
$\rho_0$ versus $m$ for $B = 21.2~\mu$T. The intersection of two
linear fits defines $m_{2-3}$, the critical magnetization. The
main figure shows $m_{2-3}$ versus $B$. The solid line is the
prediction from the SMA. } \label{gs}
\end{figure}

At long times, after the oscillations have damped out, we can
study the mean-field ground state of this system. Within the SMA,
a quantum phase transition from a three- to a two-component BEC is
predicted for the mean-field ground state~\cite{gs}. We use SG-AI
as a direct way to measure the equilibrium populations and to
observe this phase transition.

The behavior of the variance of $\rho_0$ allows us to determine
when the system has settled down to a steady state. At a given
time we measure $\rho_0$ 25-30 times and calculate a variance. In
the steady state, the variance reaches a minimum. Due to technical
noise in atom counting, this variance is larger than that
predicted by a quantum calculation of the spinor ground
state~\cite{youQ}. We find that a steady state for $m=0$ is
reached within our maximum hold time of 10~s for $ B\geq18~ \mu$T.
For non-zero $m$, the steady state is reached within a much
shorter time.

Figure~\ref{gs} (a) shows the steady state values of $\rho_0$ as a
function of $B$. For each measurement, we prepare nearly identical
initial states at $B = 24.4~\mu$T to set the populations and
magnetizations. We then wait 4 s to reach the steady state before
adiabatically ramping the field to a desired final value over 500
ms. We then wait another 5s before making an SG-AI measurement. In
the inset of Fig.~\ref{gs} (b), however, each initial state is
directly prepared with a different magnetization. In both figures,
a transition between a two-component BEC with $\rho_0=0$ and a
three-component BEC with non-zero $\rho_0$, is observed. A
critical magnetic field $B_{2-3}$ and a critical magnetization
$m_{2-3}$ are defined and extracted from the intersection of two
linear fits to the data. Good agreement is found between the
experimental value of $m_{2-3}$ and the prediction from the
SMA~\cite{gs}, as shown in Fig.~\ref{gs}(b). This confirms a phase
transition from a three- to a two-component spinor BEC in the
antiferromagnetic mean-field ground state. Although it is observed
here at a finite temperature the phase transition should remain at
zero temperature, where the transition would be driven solely by
quantum fluctuations. It has been predicted that this phase
transition persists even when the three spin states do not share
the same spatial distribution and the application of the SMA is no
longer appropriate~\cite{gs}.

In conclusion, we have demonstrated that Faraday rotation
spectroscopy provides a method to continuously monitor the spin
dynamics in an antiferromagnetic spinor BEC. The technique
provides a sharp signature to locate the magnetically tuned
boundary in phase space between the oscillating and running phase
solutions. We observe a dependence of the position of this
separatrix on the magnetization. In addition, we have confirmed a
quantum phase transition from a two- to a three-component BEC in
the mean-field ground state. We are presently investigating the
dissipation mechanism leading to the equilibrium observed in these
experiments. We think that physics beyond the SMA, possibly
similar to Landau damping of excitations of a single component
BEC~\cite{landau}, may explain the dissipation.

We thank W. D. Phillips, J. V. Porto and E. Gomez for insightful
discussions, and ONR for financial support. SEM thanks the
NIST/NRC postdoctoral program.


\end{document}